\documentclass[]{tPHM2e}
\begin{document}

\newcommand{\bols}{\mathbf{s}}
\newcommand{\bJ}{\mathbf{J}}
\newcommand{\bC}{\mathbf{C}}
\newcommand{\bD}{\mathbf{D}}
\newcommand{\bA}{\mathbf{A}}

\title{Effect of coupling asymmetry on mean-field solutions of direct and inverse Sherrington-Kirkpatrick model}

\author{Jason Sakellariou$^{\rm a}$, Yasser Roudi$^{\rm bc}$,Marc Mezard $^{\rm a}$,John Hertz$^{\rm cd}$\\
\vspace{6pt}  
$^{\rm a}${\em{LPTMS, CNRS and Universit\'e Paris-Sud, 91405 Orsay Cedex, France}};
$^{\rm b}${\em{Kavli Institute for Systems Neuroscience, NTNU, 7014 Trondheim, Norway}};
$^{\rm c}${\em{NORDITA, 10691 Stockholm, Sweden}}; 
$^{\rm d}${\em{The Niels Bohr Institute, , 2100 Copenhagen, Denmark}}
}

\maketitle

\begin{abstract}
We study how the degree of symmetry in the couplings influences the performance of
three mean field methods used for solving the direct and inverse problems for generalized
Sherrington-Kirkpatrick models. In this context, the direct problem is predicting the potentially
time-varying magnetizations. The three theories include the first and second order Plefka expansions, 
referred to as naive mean field (nMF) and TAP, respectively, and a mean field theory which is exact 
for fully asymmetric couplings. We call the last of these simply MF theory. We show that for the
direct problem, nMF performs worse than the other two approximations, TAP outperforms MF when 
the coupling matrix is nearly symmetric, while MF works better when it is strongly asymmetric. 
For the inverse problem, MF performs better than both TAP and nMF, although an ad hoc adjustment 
of TAP can make it comparable to MF.  For high temperatures the performance of TAP and MF approach 
each other.
\bigskip
\end{abstract}

\section{Introduction}
Predicting the dynamical properties of a disordered system given a specific realisation of its 
parameters is an old and important problem in statistical mechanics. This is what one can call 
a direct problem. Apart from being important on its own, solving the direct problem is also a 
crucial step in solving the inverse problem: inferring the parameters of a system from 
measurements of its dynamics. With the rapid advance of methods for observing the dynamics 
of biological systems composed of many elements, the inverse problem has received a lot of 
recent attention. This line of research has allowed inferring functional and physical connections 
in neuronal networks \cite{Schneidman05,Shlens06,Cocco09,Roudi09,Roudi11}, gene regulatory 
networks \cite{Lezonetal06} and protein residue contacts \cite{WeigtPNAS2009}.

A useful platform for studying the inverse problem is a dynamical version of the 
Sherrington-Kirkpatrick (SK) model: a set of $N$ classical spins, $s_i=\pm 1$ subject to 
a potentially time-varying external field $h_i(t)$ with couplings $J_{ij}$ between 
them  and a stochastic update rule. In the direct problem one tries to predict the
magnetizations $m_i(t)$ given the coupling and fields. 
In the inverse problem one does the opposite, i.e. one infers the couplings and the
fields from measured magnetizations and correlations. 

When the system is in equilibrium and the distribution of states
follows the Boltzmann distribution, several approaches for both direct
and inverse problems have been developed. These include both exact and
approximate iterative algorithms, such as Boltzmann learning and
Susceptibility propagation \cite{AurellOllionRoudi10,MezardMora}
relating the magnetizations to model parameters, as well as
closed-form equations based on naive mean field (nMF) and TAP
\cite{Tanaka98,Kappen98} equations for the SK model. Motivated by the
fact that biological systems are usually out of equilibrium, some
recent work has focused on reconstructing the parameters of a
dynamical  Ising spin glass model obeying either synchronous or
asynchronous updating from observing its out-of-equilibrium dynamics \cite{Roudi11,Zeng11,MezardJstat11}.

In this paper, we investigate how three recently proposed mean field
methods for the direct and inverse problems perform on models with
different degrees of symmetry in their coupling matrices. The three
methods are the nMF and TAP equations, derived using the
high-temperature Plefka expansions  of the generating functional to
first order and second order  \cite{Roudi11-2}, and a mean field theory (denoted simply MF) \cite{MezardJstat11} that
is exact for the SK model with
fully asymmetric couplings.

\section{Solutions to the direct and inverse problems}

We consider a model in which the probability of being in state $\bols$ at time step $t$, $p_t(\bols)$, is given by 
\begin{subequations}
\begin{equation}
p_{t+1} (\bols)=\sum_{\bols'} W_t[\bols;\bols'] p_{t}(\bols')\label{DynSK-syna}
\end{equation}
\begin{equation}
W_t[\bols;\bols']=\prod_i \frac{\exp(s_i \theta_i)}{2\cosh \theta_i }\label{DynSK-synb}
\end{equation}
\begin{equation}
\theta_i(t)=h_i(t)+\sum_j J_{ij} s'_j(t)\label{DynSK-sync}.
\end{equation}
\label{DynSK}
\end{subequations}
For the choice of couplings $J_{ij}$, we follow \cite{Crisanti87}, taking 
\begin{equation}
J_{ij}=J^{sym}_{ij}+ k J^{asym}_{ij}\label{Js}
\end{equation}
where $J^{sym}_{ij}=J^{sym}_{ji}$ is the symmetric part of the couplings while $J^{asym}_{ij}=-J^{asym}_{ji}$
is the antiymmetric part. All the couplings $J^{sym}_{ij}$ and $J^{asym}_{ij}$ are drawn independently from a
zero-mean Gaussian distribution with variance 
\begin{equation}
\overline{[J^{symm}_{ij}]^2}=\overline{[J^{asym}_{ij}]^2}=\frac{g^2}{(1+k^2)N}. \label{varJ}
\end{equation}
With Eqs. \ref{Js} and \ref{varJ}, the couplings $J_{ij}$ have variance of $g^2/N$ and the degree of symmetry is controlled
by $k$: for $k=0$ the model is fully symmetric ($J_{ij}=J_{ji}$) while for $k=1$, it is fully asymmetric ($J_{ij}$ independent of $J_{ji}$).

The direct problem consists in estimating the instantaneous
magnetization of spin $i$ at time $t$, $m_i(t)$. The estimation
obtained from the nMF, TAP and MF are respectively:
\begin{subequations}
\begin{equation}
m_i(t+1)=\tanh\Big[h_i(t)+\sum_j J_{ij} m_j(t)\Big] \label{nMF}
\end{equation}
\begin{equation}
m_i(t+1)=\tanh\Big[h_i(t)+\sum_j J_{ij} m_j(t)-m_i(t+1) \sum_j J^2_{ij} (1-m_j^2(t))\Big] \label{TAP}
\end{equation}
\begin{equation}
m_i(t+1)=\int \frac{dx}{\sqrt{2\pi}} e^{-x^2/2} \tanh\Big[h_i(t)+\sum_j J_{ij} m_j(t)+x\sqrt{\Delta_i(t)}\Big] \label{MF}
\end{equation}
\label{deirectEq}
\end{subequations}
where in the last equation
\begin{equation}
\Delta_i(t) = \sum_j J_{ij}^2(1-m_i^2(t))\ .
\label{Deltadef}
\end{equation}

For deriving Eqs. \ref{nMF} and \ref{TAP}, i.e. nMF and TAP, one first writes down the generating 
functional for the process defined by Eq.\ \ref{DynSK}, performs a Legendre transform to fix the 
magnetizations and expands the results for small $g$ (i.e. high temperature). To the first order, 
this expansion gives the nMF equations, Eq.\ \ref{nMF}. Keeping terms up to the second order yields 
a correction to the nMF equations resulting in the the TAP equations, Eq. \ref{TAP},
for this dynamical model. nMF and TAP are, therefore, high temperature
expansions for an arbitrary set of couplings, with no assumption about
their distribution or its degree of symmetry. The third equation is
derived for arbitrary $g$, but under the mean-field assumption that at
each time step the fields acting on the spins are independent Gaussian
variables. This is exact for this SK model 
when the coupling matrix is fully asymmetric i.e. when $k=1$.

These direct equations can also be used for solving the inverse
problem. The idea is to use the data in order to measure the
magnetizations $m_i(t)$, the equal time correlations  $C_{ij}=\langle
\delta s_i(t) \delta s_j(t) \rangle$, and the time-delayed
correlations $ D_{ij}=\langle \delta s_i(t+1) \delta s_j(t)\rangle$,
where $\delta s_i(t)=s_i(t)-m_i(t)$. For
the process in Eq.\ \ref{DynSK}, one can write the time-delayed correlations as
\begin{equation}
D_{ij}=\langle \tanh\big[\theta_i(t)\big] s_j(t)\rangle-\langle \tanh\big[\theta_i(t)\big] \rangle \langle s_j(t)\rangle.
\label{D}
\end{equation}
To derive the inverse TAP and nMF, one then uses Eq.\ \ref{D}, expands the $\tanh$ around $m_i$ that 
satisfies one of the direct equations \ref{nMF} and \ref{TAP}. In the case of
MF, one writes an expression for the joint distribution of $\theta_i(t)$ and $\theta_j(t)$
that is exact for a fully asymmetric SK model. This joint distribution can then be used to 
relate $\bJ \bD$ to $\bC$ in the limit of small $C_{ij}$; for details see \cite{Roudi11,MezardJstat11}.
Within all three approximations, nMF, TAP, and MF, the resulting expression takes the form
\begin{equation}
\bD=\bA \bJ \bC\ ,
\label{DAJC}
\end{equation}
where the matrix $A$ is a diagonal matrix that depends on the
approximation:
\begin{subequations}
\begin{equation}
A^{\rm nMF}_{ij}=\delta_{ij} (1-m_i^2) 	\ ,	\label{AnMF}
\end{equation}
\begin{equation}
A^{\rm TAP}_{ij}=\delta_{ij} (1-m_i^2)(1-F_i)\ ,		\label{ATAP}
\end{equation}
\begin{equation}
A^{\rm MF}_{ij}=\delta_{ij} \int  \frac{dx}{\sqrt{2\pi}} e^{-x^2/2}
  \Big[1-\tanh^2 (h_i(t)+\sum_j J_{ij} m_j+x\sqrt{\Delta_i}) \Big]\ .\label{AMF}
\end{equation}
\end{subequations}
In Eq.\ \ref{ATAP} $F_i$ satisfies a cubic equation. For details see \cite{Roudi11} 
and \cite{MezardJstat11}. Not surprisingly, expanding Eq.\ \ref{AMF} to linear
or second order in $J_{ij}$ yields $A^{\rm nMF}$ and $A^{\rm TAP}$ in
Eqs.\ \ref{AnMF} and \ref{ATAP}, respectively.

Eq.\ \ref{DAJC} can be solved for $\bJ=\bA^{-1} \bD \bC^{-1}$,
provided one has enough data so that the estimation of $C$ is good,
allowing its numerical inversion.

\begin{figure}[htb] 
\center
    \subfigure{\includegraphics[height=1.5in, width=2in]{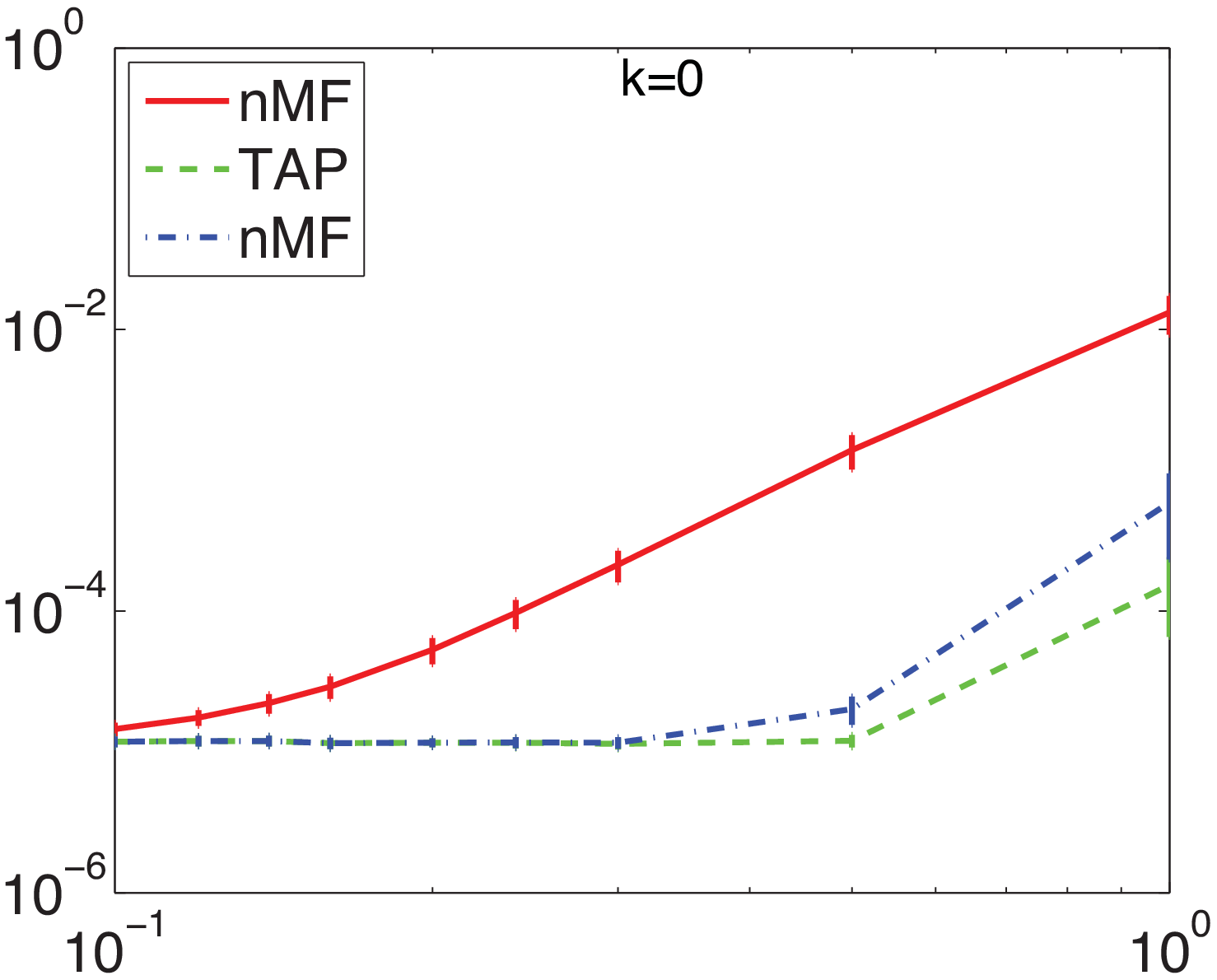}}
    \subfigure{\includegraphics[height=1.5in, width=2in]{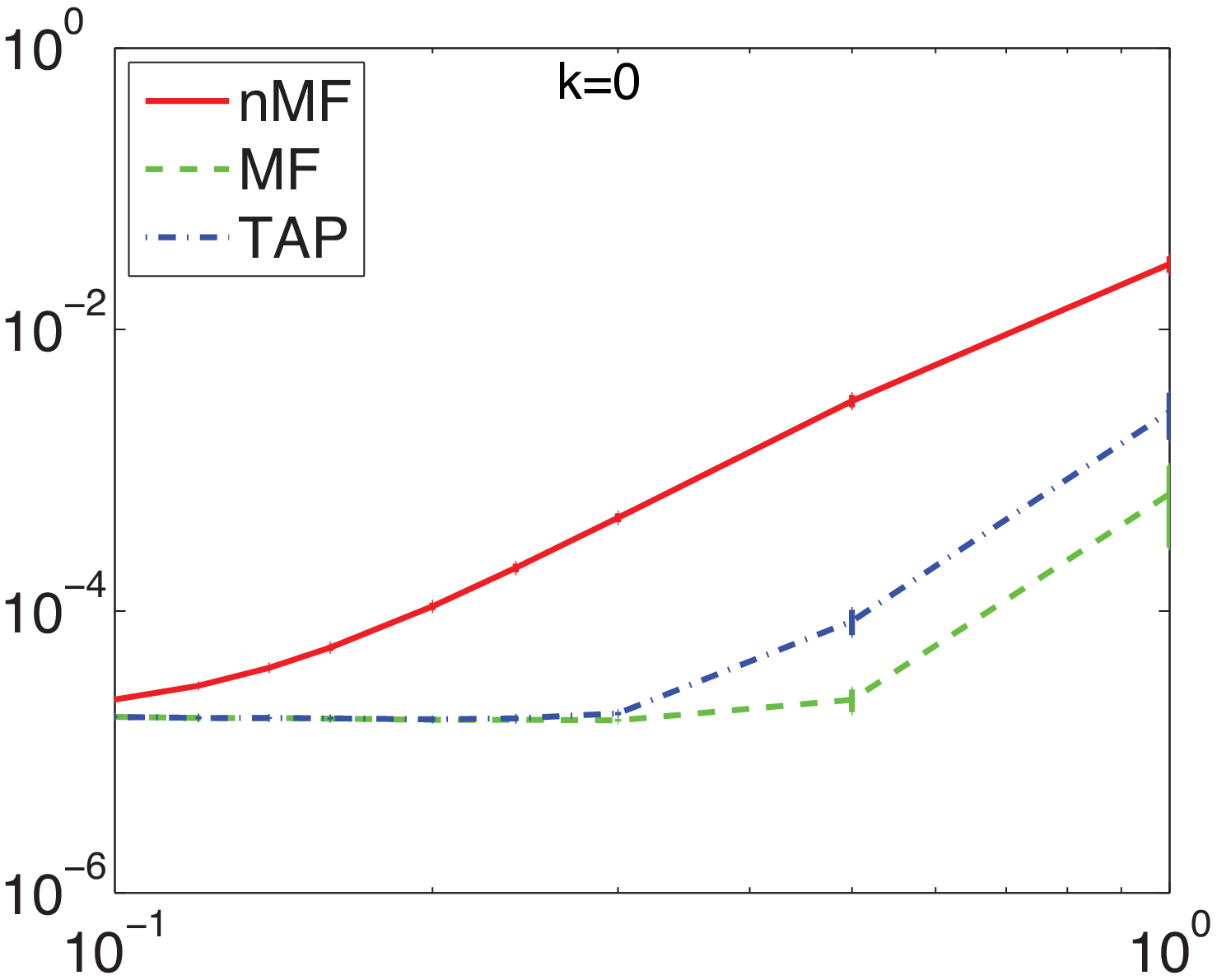}}\\
      \subfigure{\includegraphics[height=1.5in, width=2in]{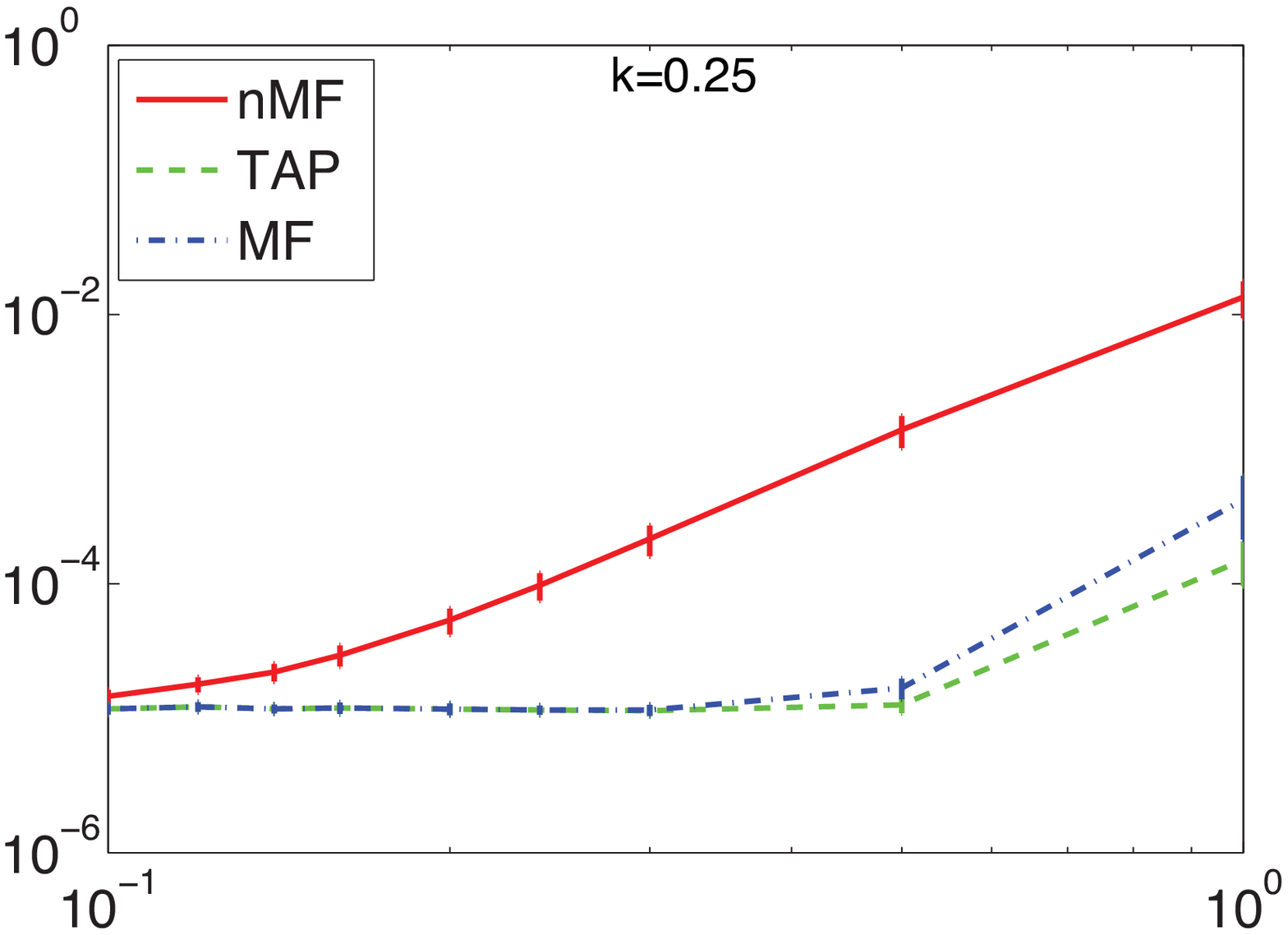}}
    \subfigure{\includegraphics[height=1.5in, width=2in]{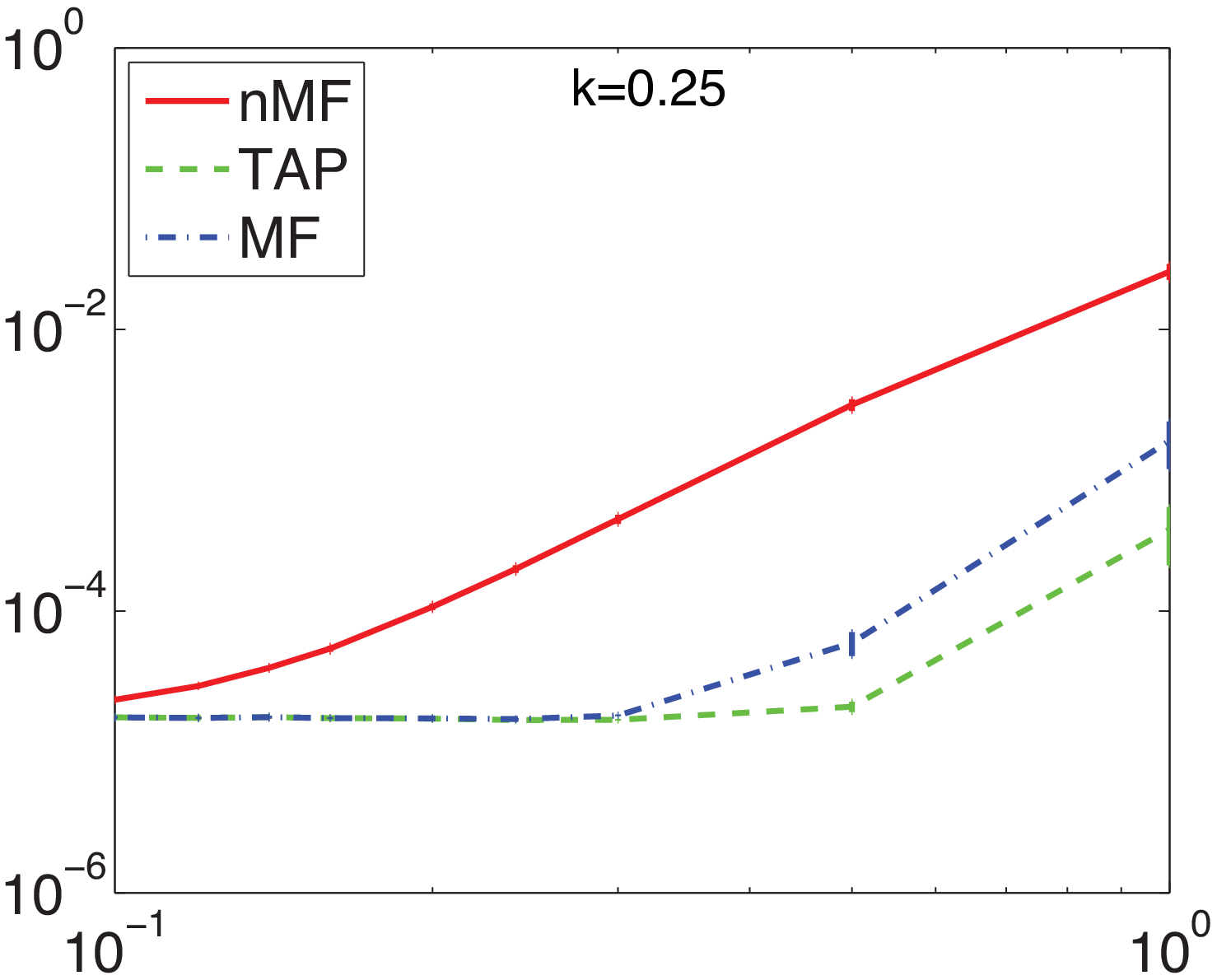}}\\
      \subfigure{\includegraphics[height=1.5in, width=2in]{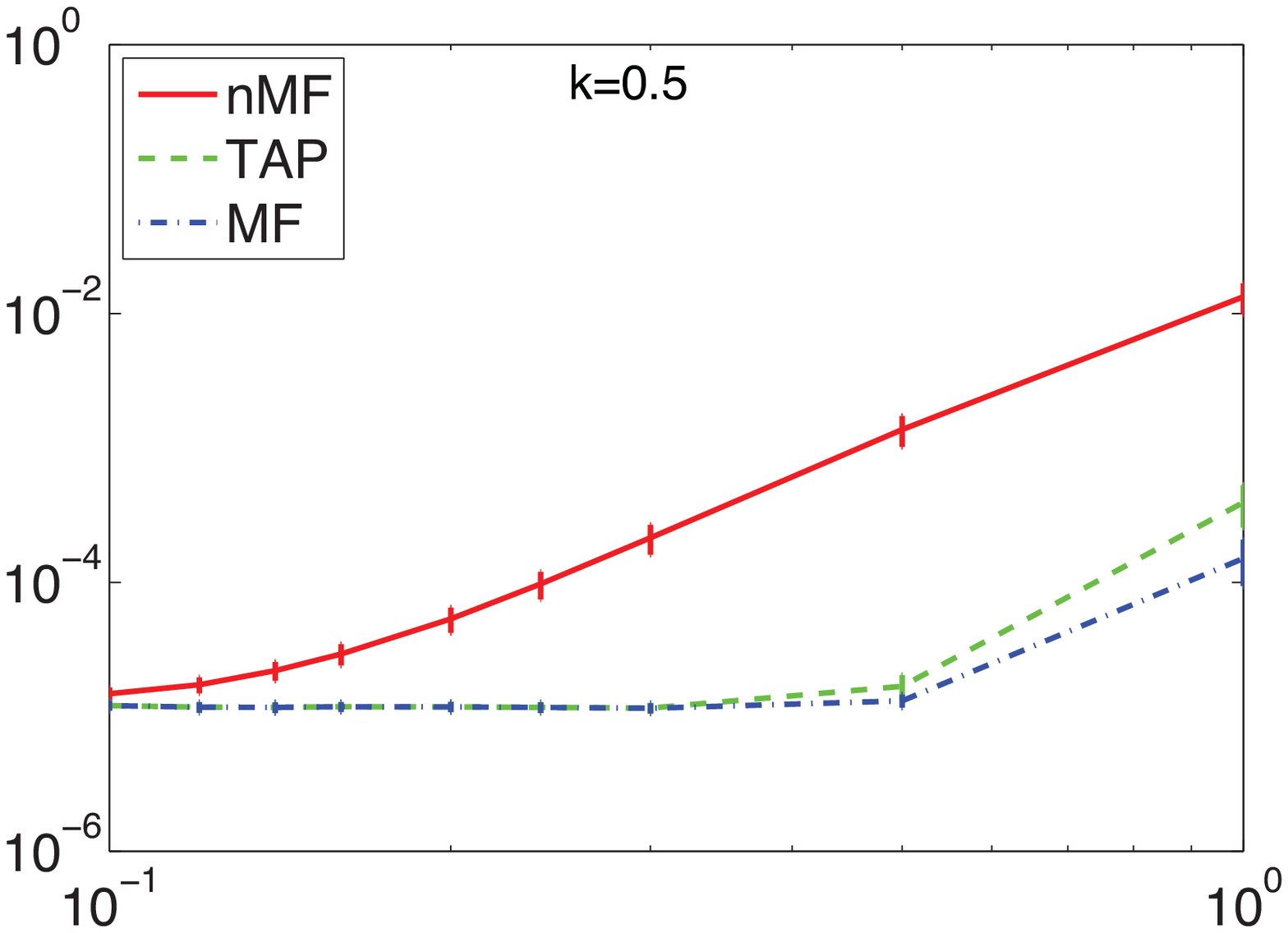}}
    \subfigure{\includegraphics[height=1.5in, width=2in]{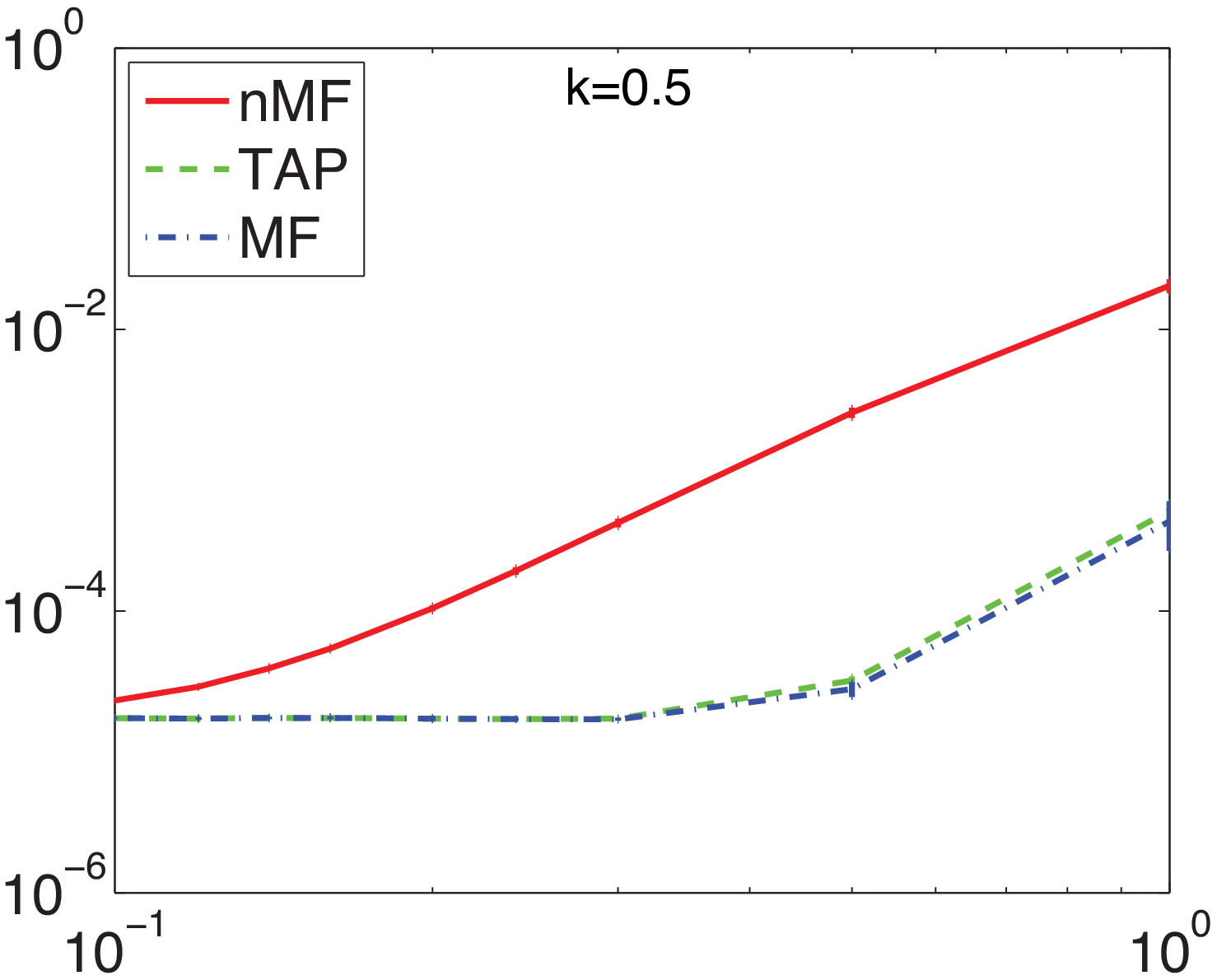}}\\
      \subfigure{\includegraphics[height=1.5in, width=2in]{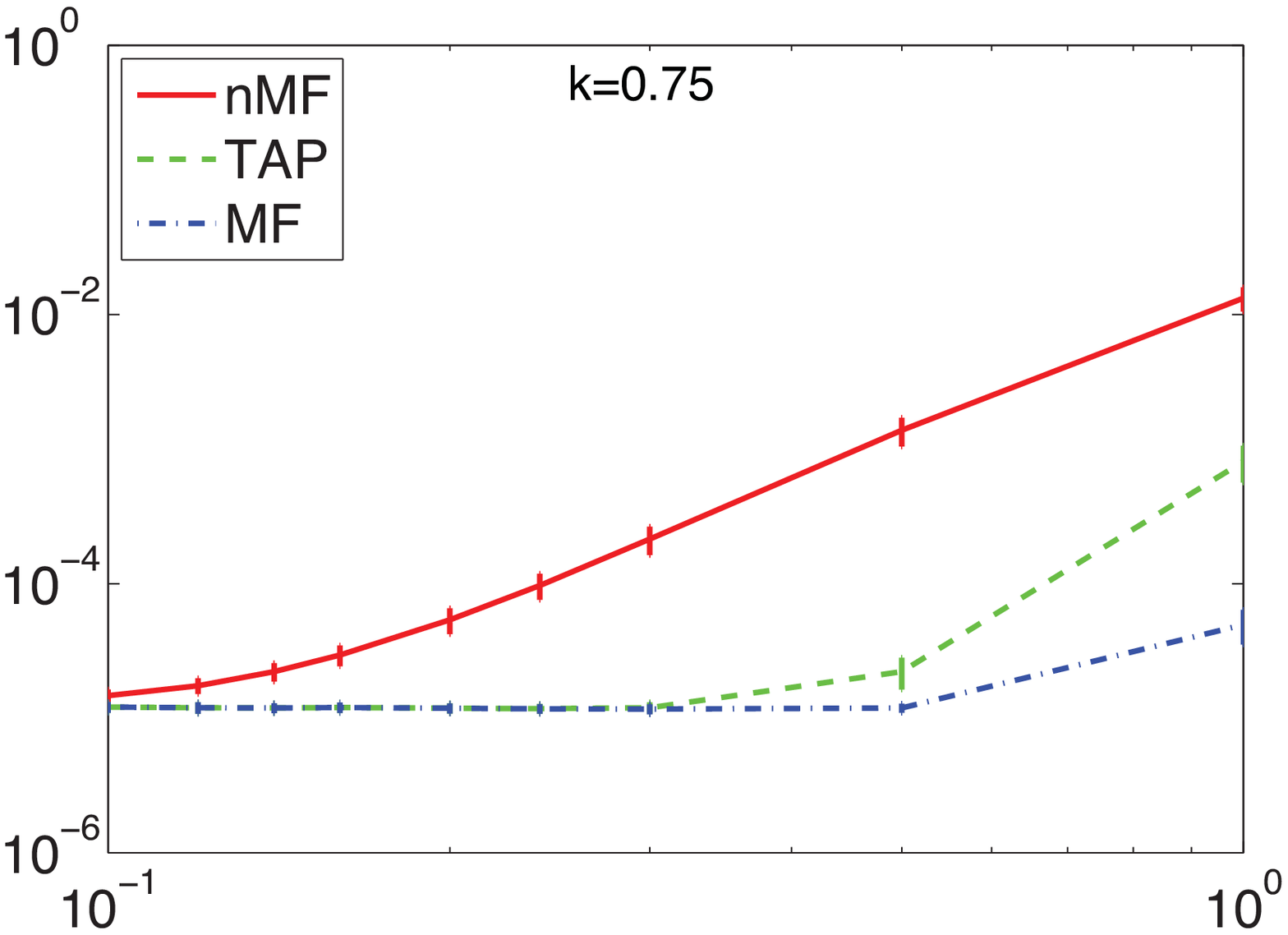}}
    \subfigure{\includegraphics[height=1.5in, width=2in]{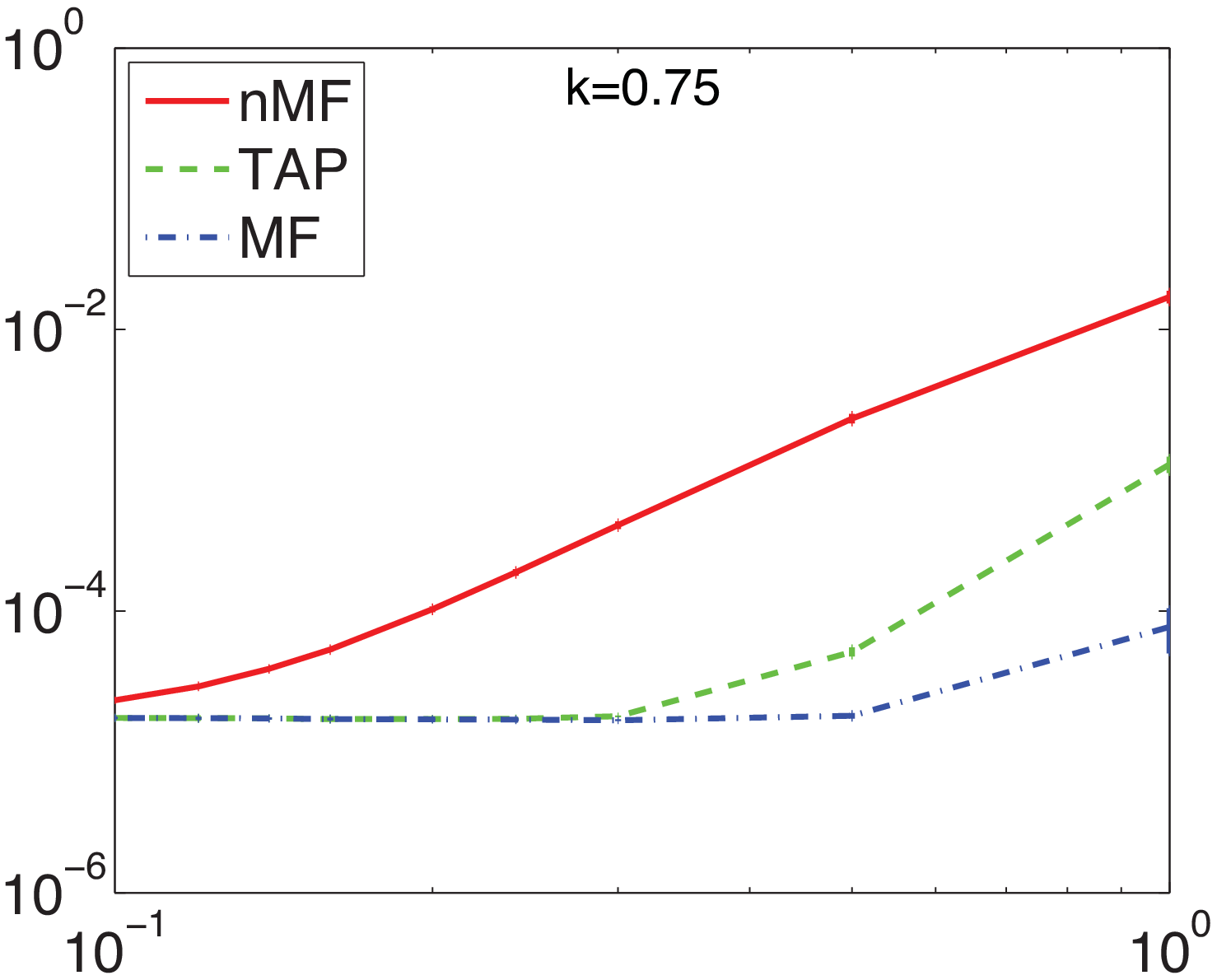}}\\
      \subfigure{\includegraphics[height=1.5in, width=2in]{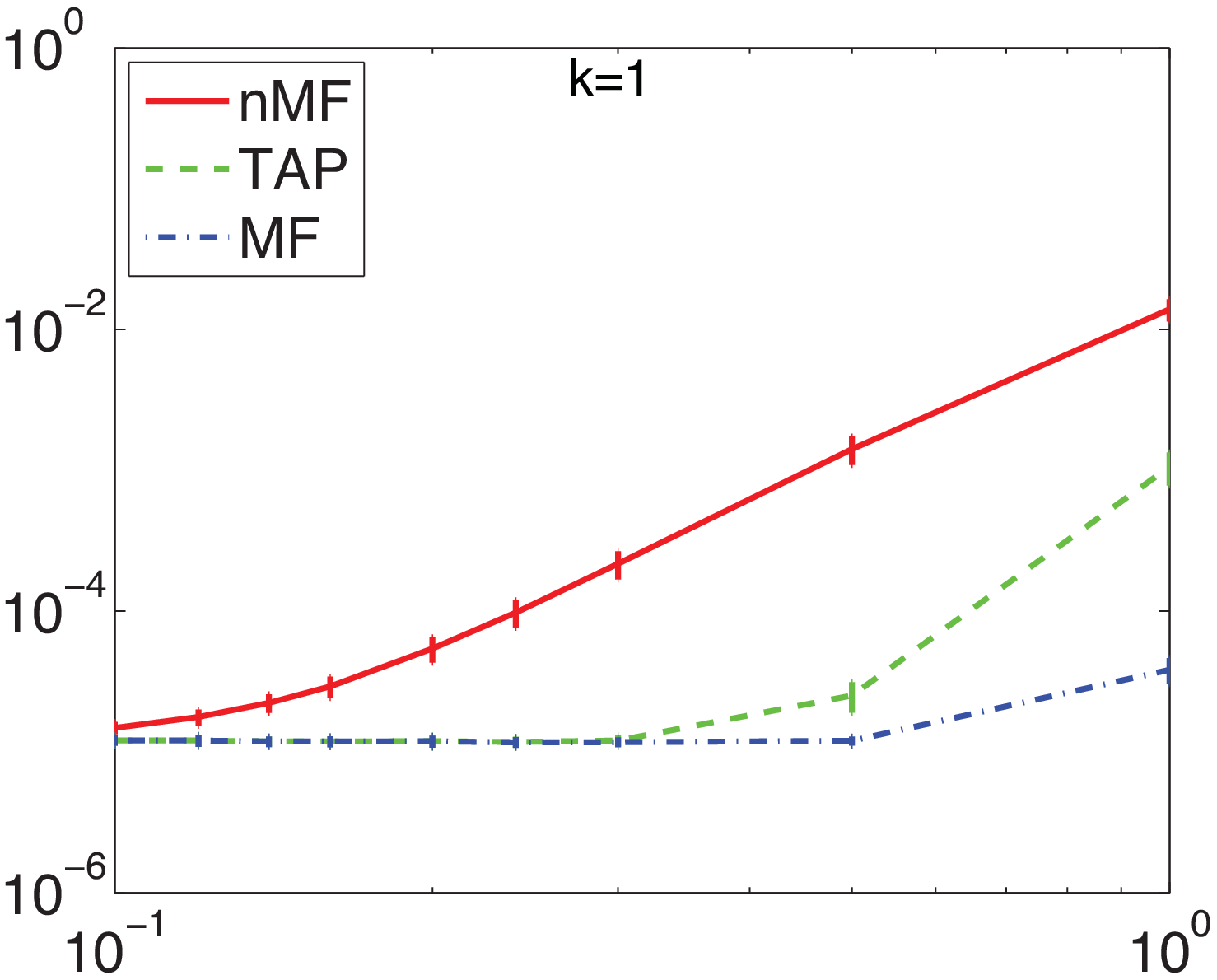}}
    \subfigure{\includegraphics[height=1.5in, width=2in]{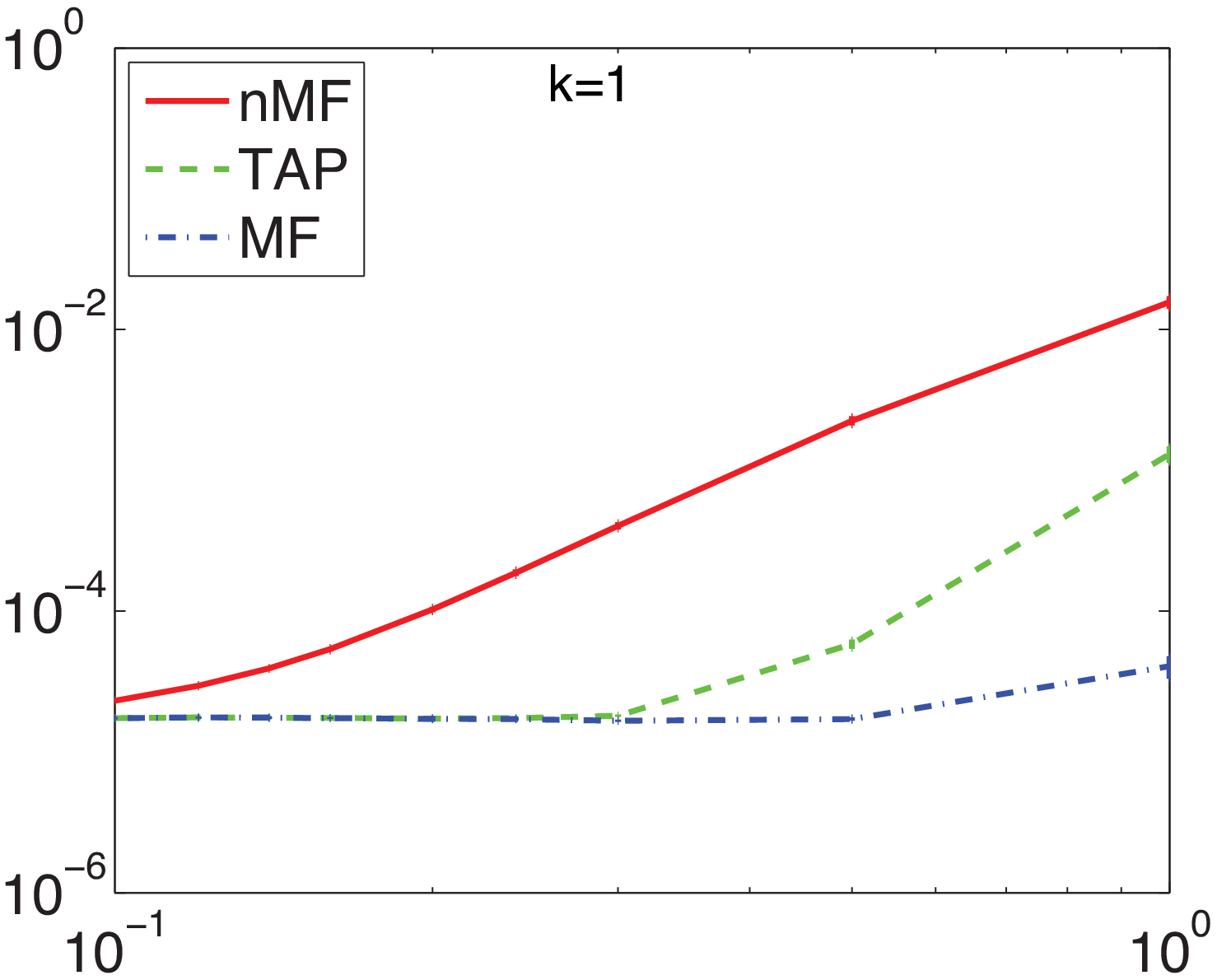}}
   \caption{

LEFT PANELS: Mean squared error of the three methods for predicting the magnetizations
at time $t$ given at time $t-1$, averaged over spins and times,
     $\overline{( m_i^{\text{predicted}}(t) -  m_i^{\text{measured}}(t))^2}$. This mean squared error
     is plotted as a function of  
     $g$ for a system of size $N=50$ with a temporally constant field drawn independently
     for each spin from a normal distribution. We have used $100$ time steps and $50000$ repeats to calculate
     the experimental magnetizations and have averaged the errors over $10$ realizations of the 
     couplings. The different figures correspond to different values of $k$. From top to bottom $k=0,0.25,0.5,0.75,1$.
     RIGHT PANELS: The same but with the addition of a sinusoidal external field of period $10$ 
     time steps and amplitude $0.5$.
  }  
  \label{fig:figDirect}
\end{figure}

\section{Effect of Symmetry}
As mentioned before, for the direct problem, we expect that the MF
becomes exact for $k=1$ for any coupling strength $g$. TAP equations
should also become exact for $k=0$ in the limit of weak
couplings. This is shown in Fig.\ \ref{fig:figDirect}, where we plot the
mean squared error in predicting the magnetizations at time $t+1$ given
the magnetizations at time $t$. This is done both for a constant field
and for an external field that varies sinusoidally with time.
 As can be seen in this figure, for both types of external fields, TAP
 equations outperform the other two methods for small $k$. 
As temperature is increased, all three approximations perform better
and become almost equally good. 
As $k$ increases, MF wins over TAP while nMF performs worse than both of them.

\begin{figure}[htb] 
\center
    \subfigure{\includegraphics[height=1.5in, width=2in]{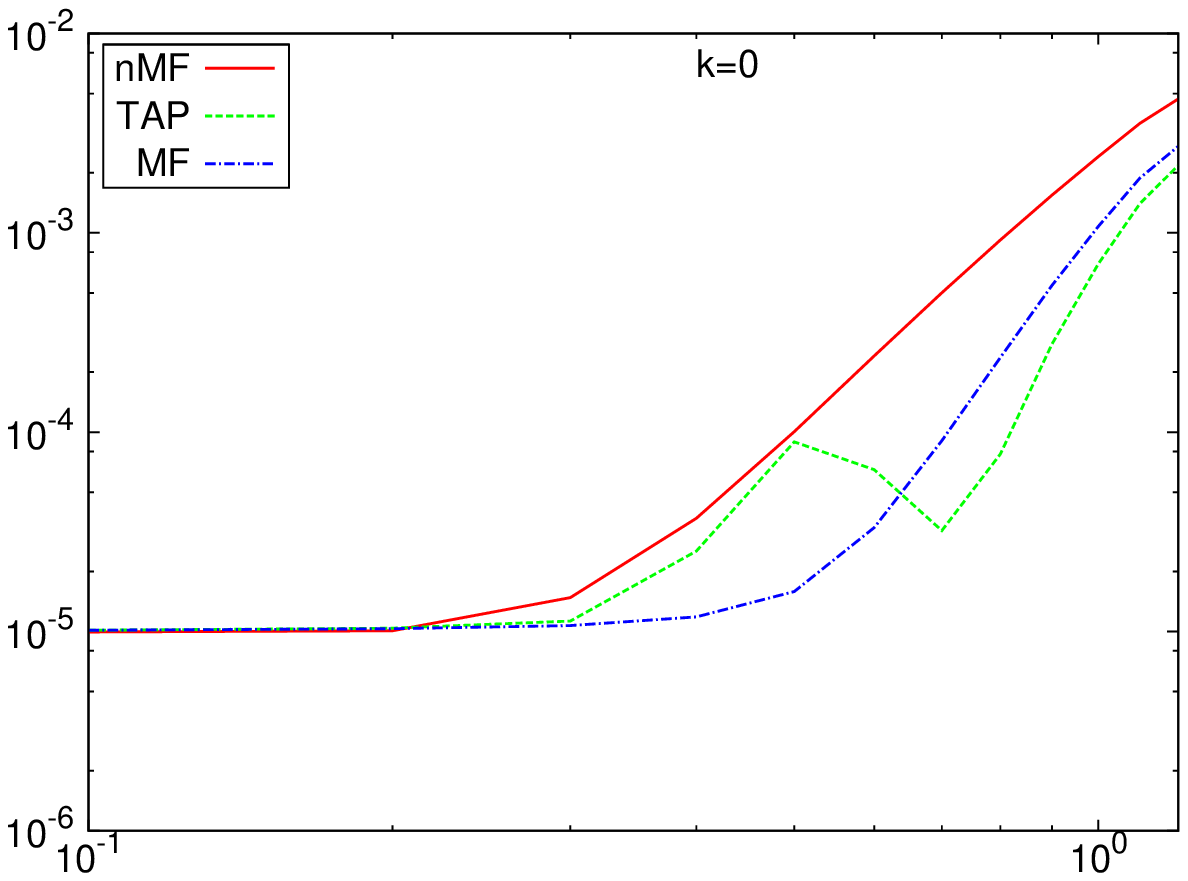}}
    \subfigure{\includegraphics[height=1.5in, width=2in]{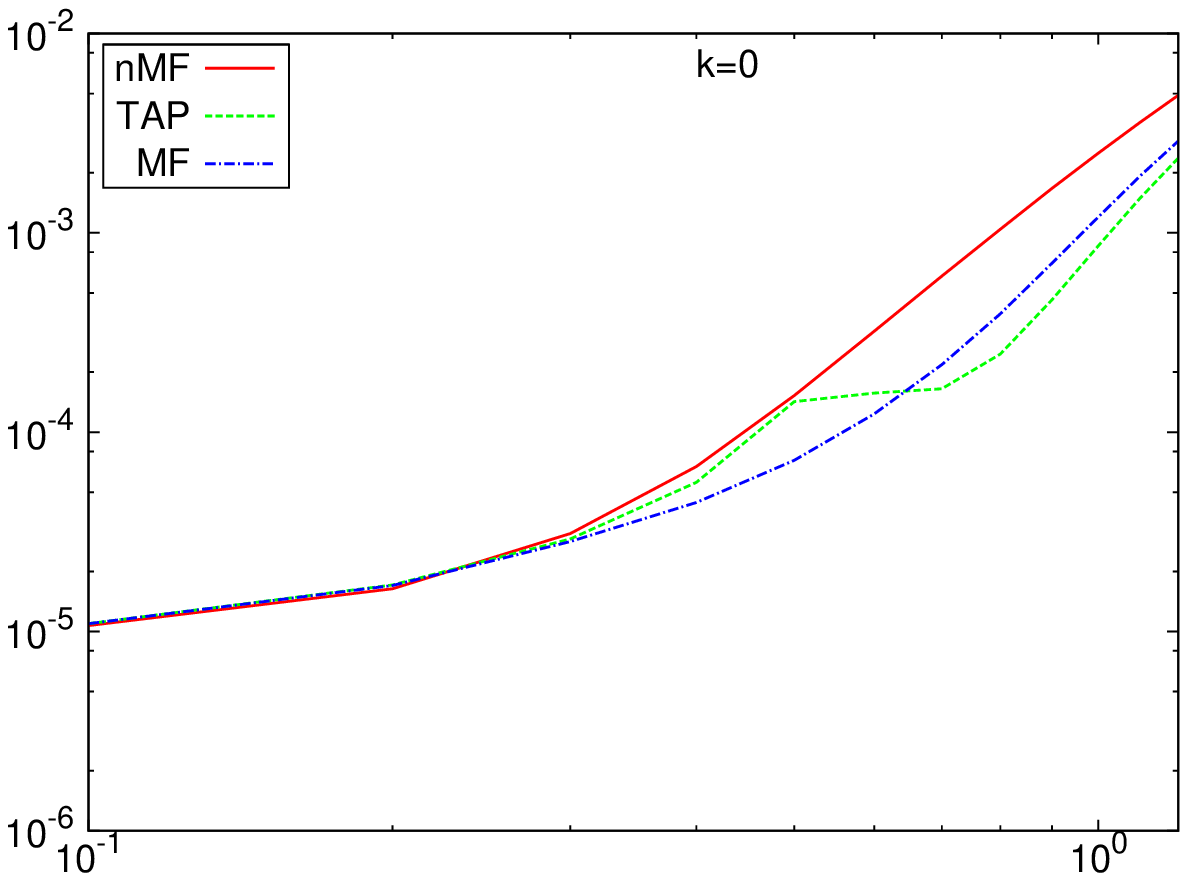}}\\
      \subfigure{\includegraphics[height=1.5in, width=2in]{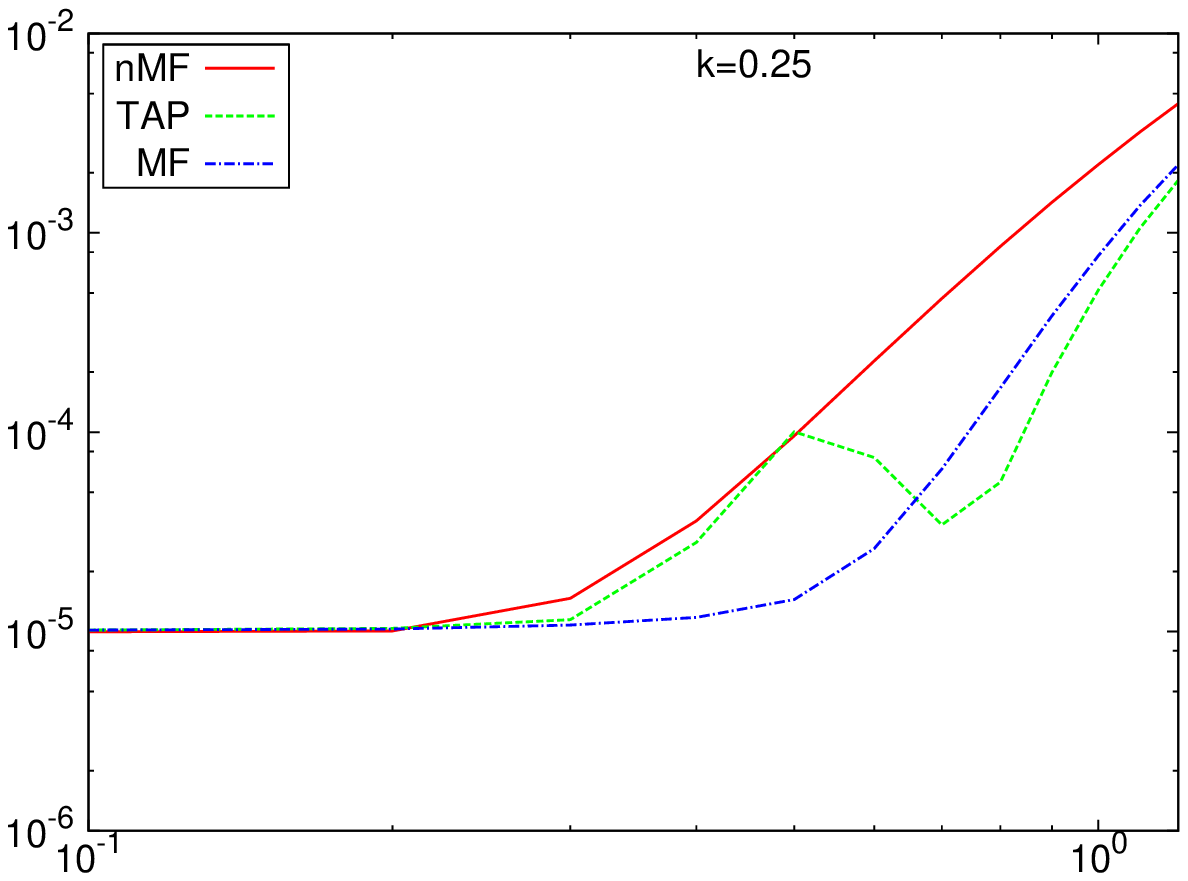}}
    \subfigure{\includegraphics[height=1.5in, width=2in]{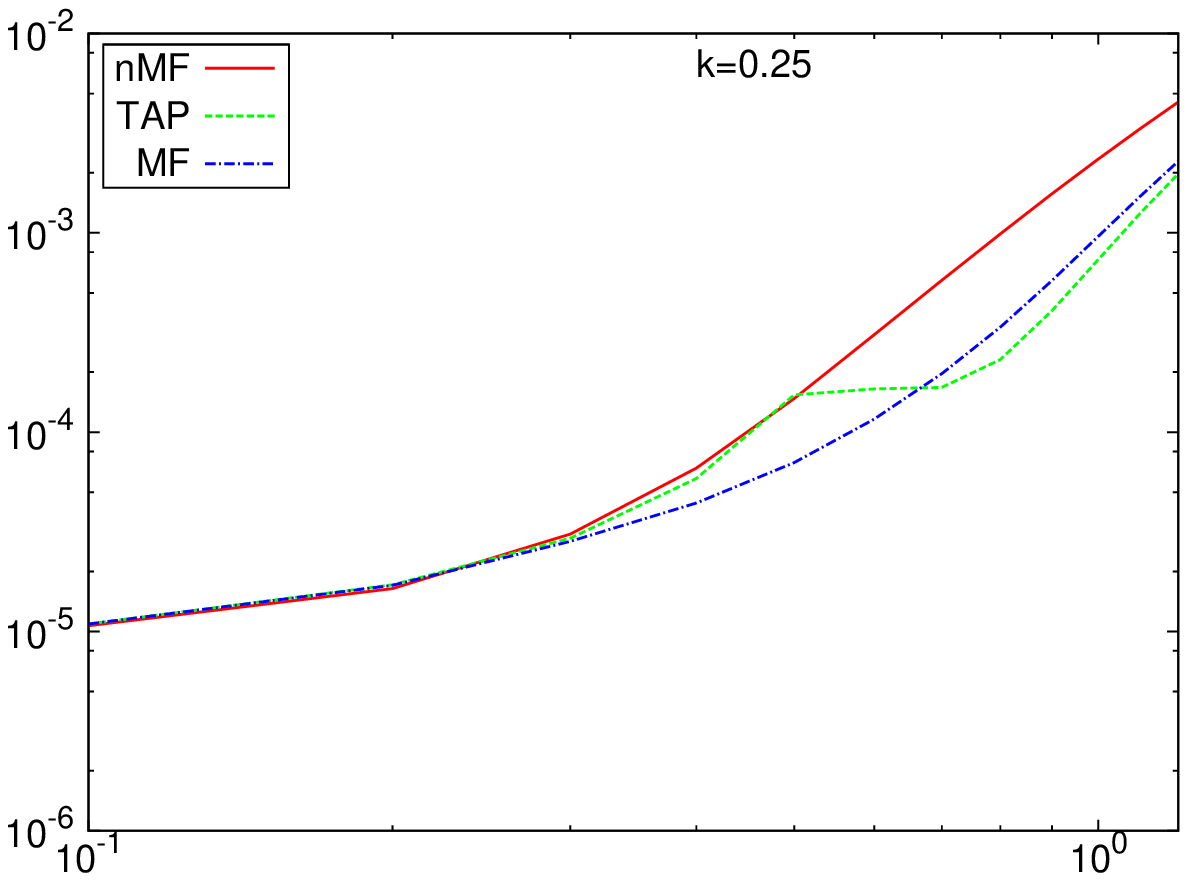}}\\
      \subfigure{\includegraphics[height=1.5in, width=2in]{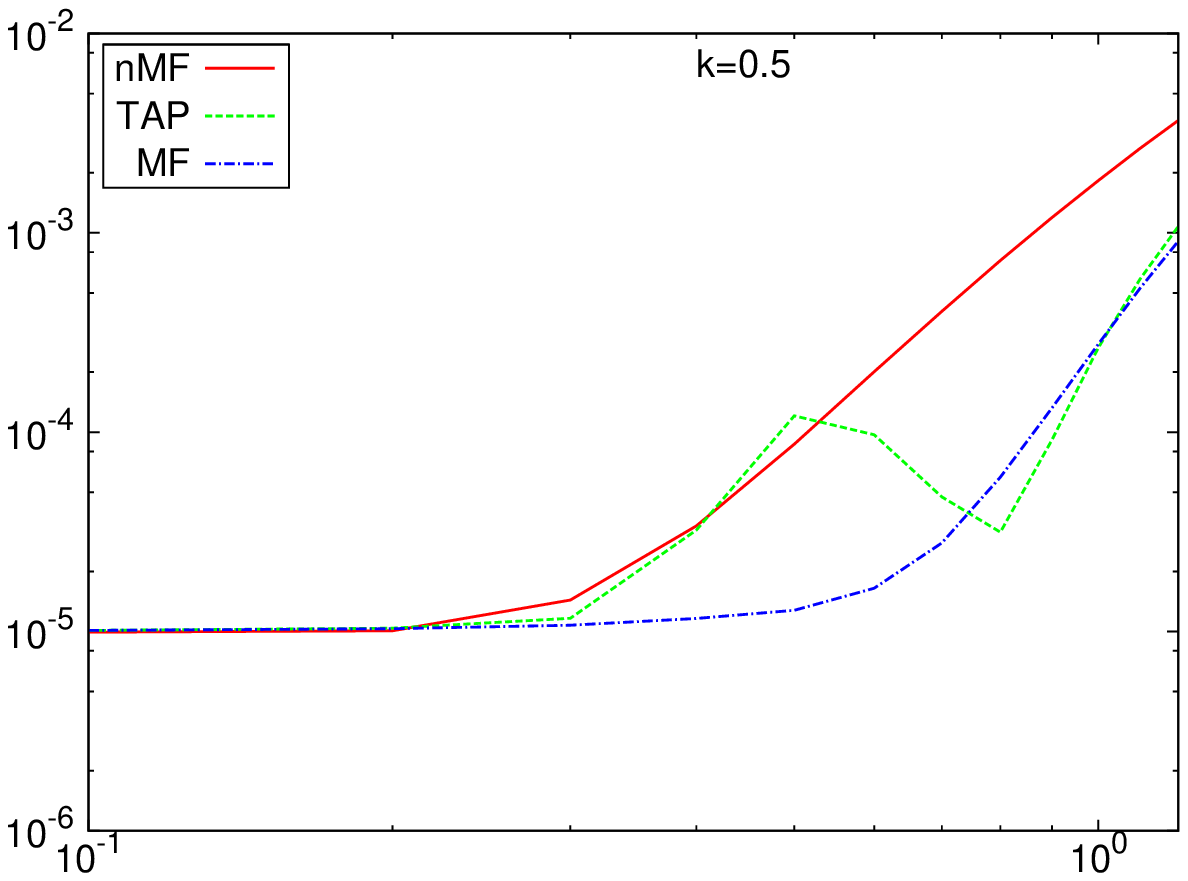}}
    \subfigure{\includegraphics[height=1.5in, width=2in]{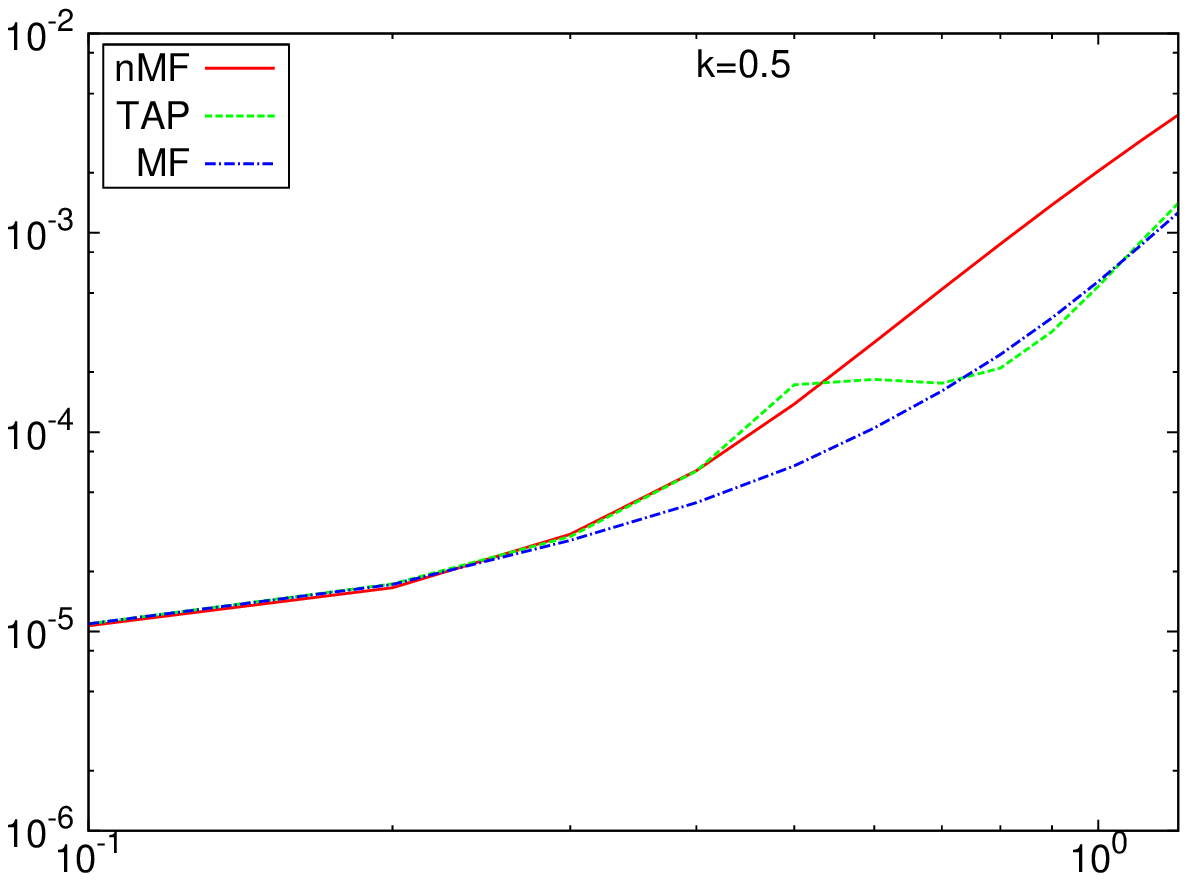}}\\
      \subfigure{\includegraphics[height=1.5in, width=2in]{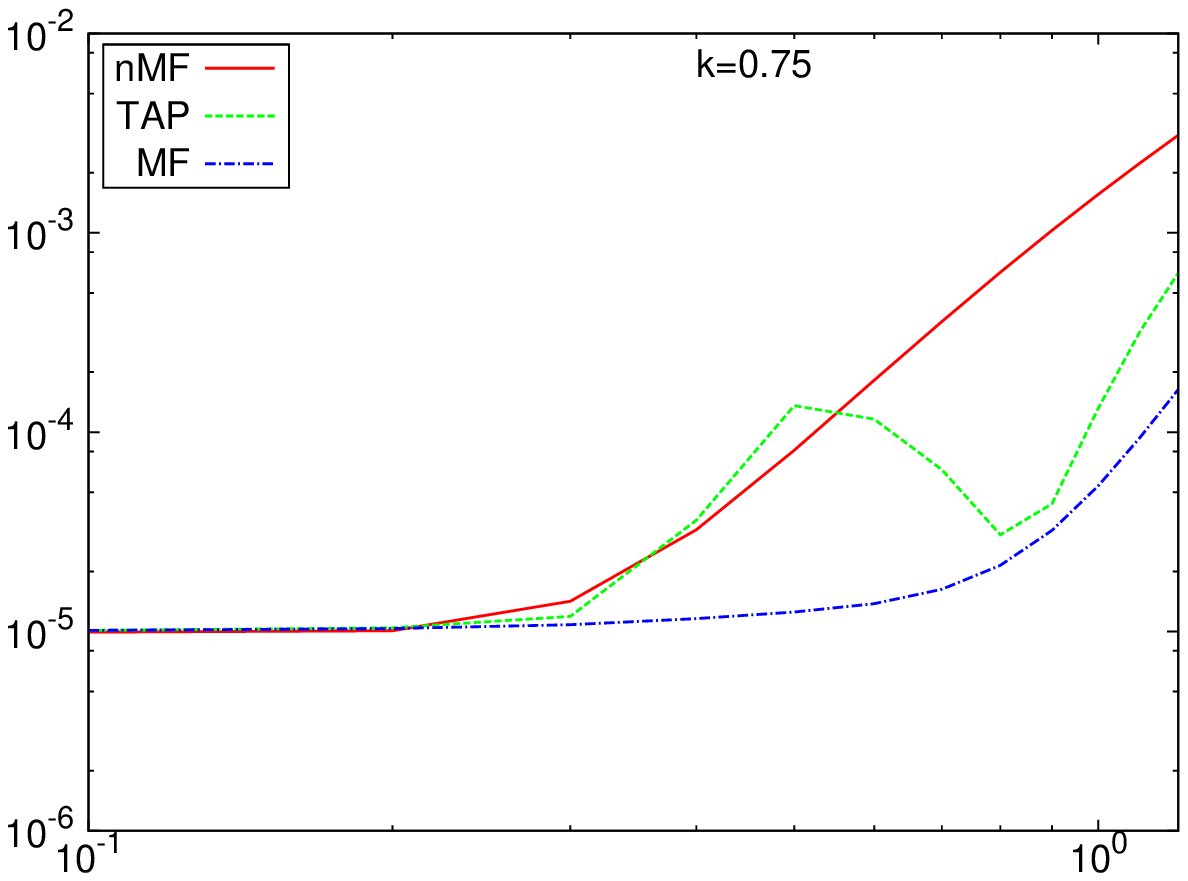}}
    \subfigure{\includegraphics[height=1.5in, width=2in]{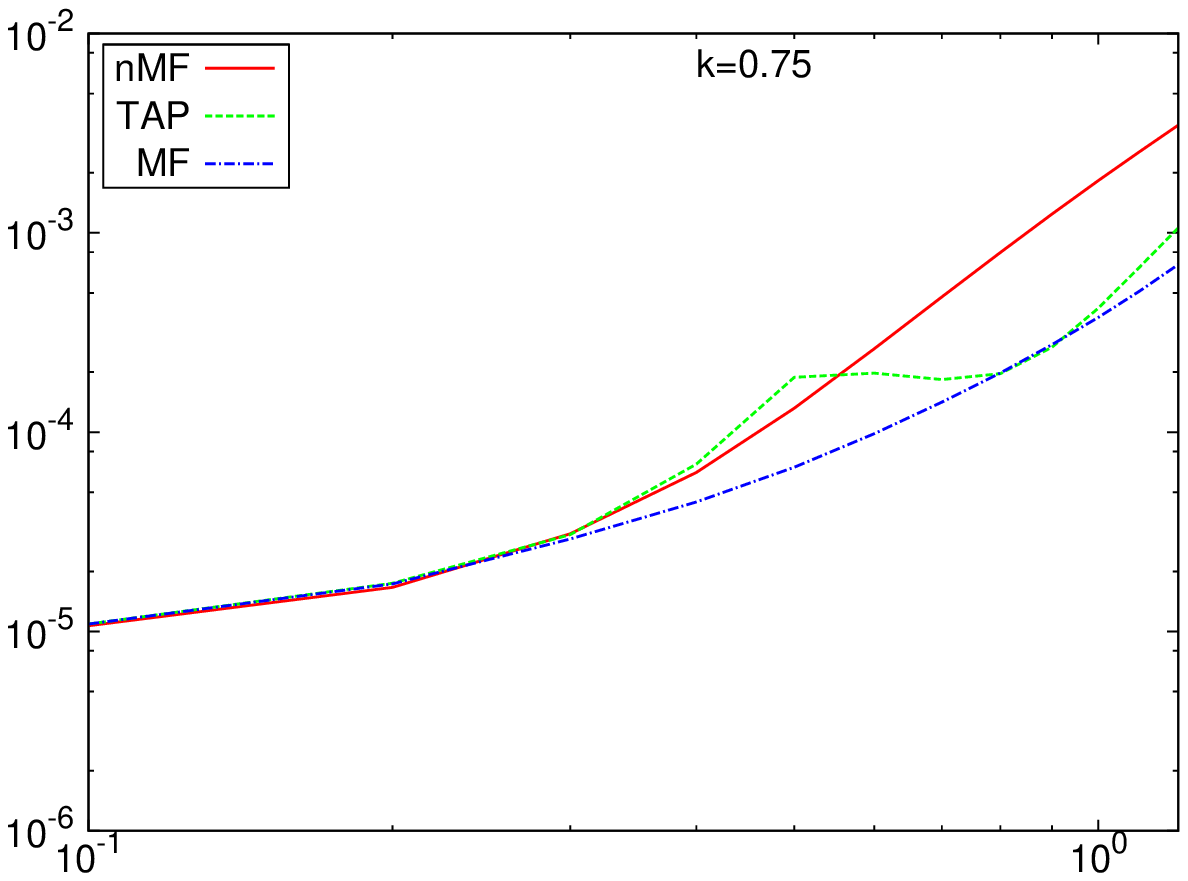}}\\
      \subfigure{\includegraphics[height=1.5in, width=2in]{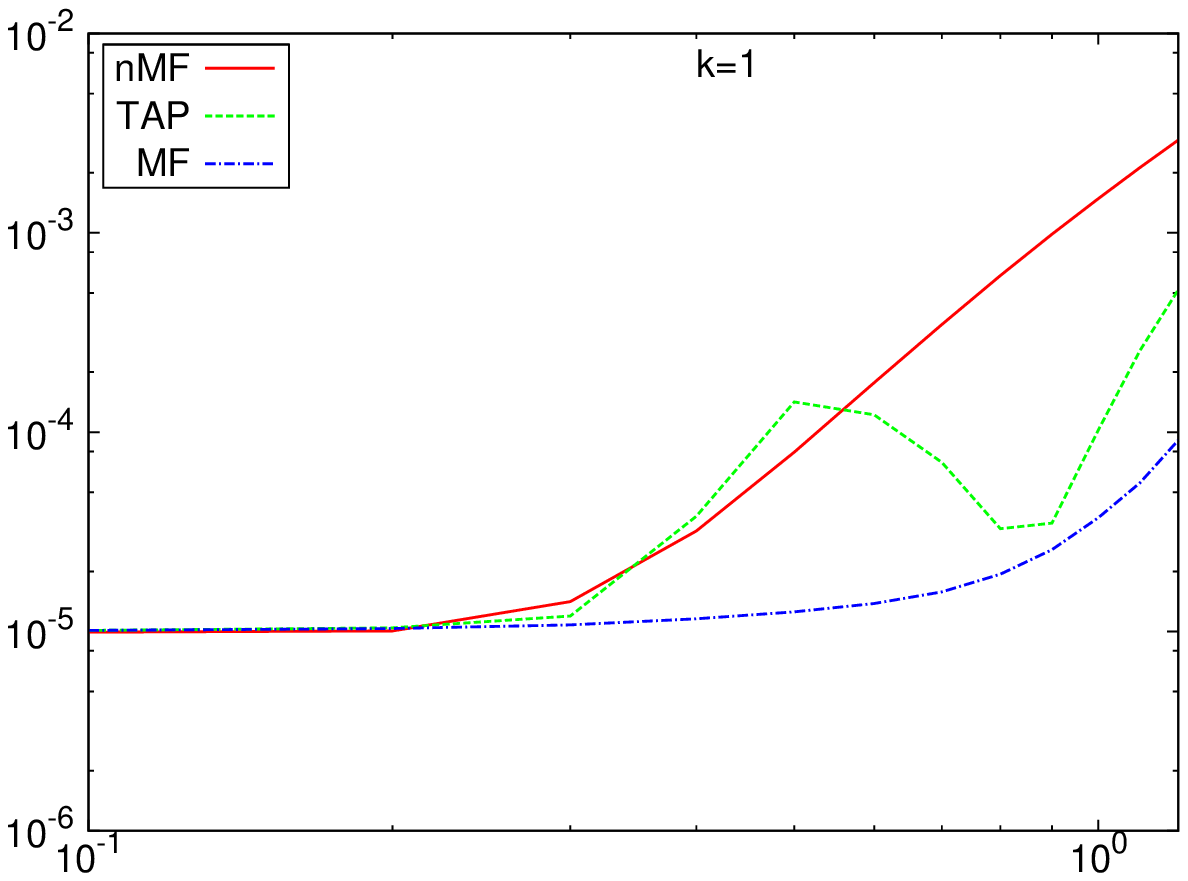}}
    \subfigure{\includegraphics[height=1.5in, width=2in]{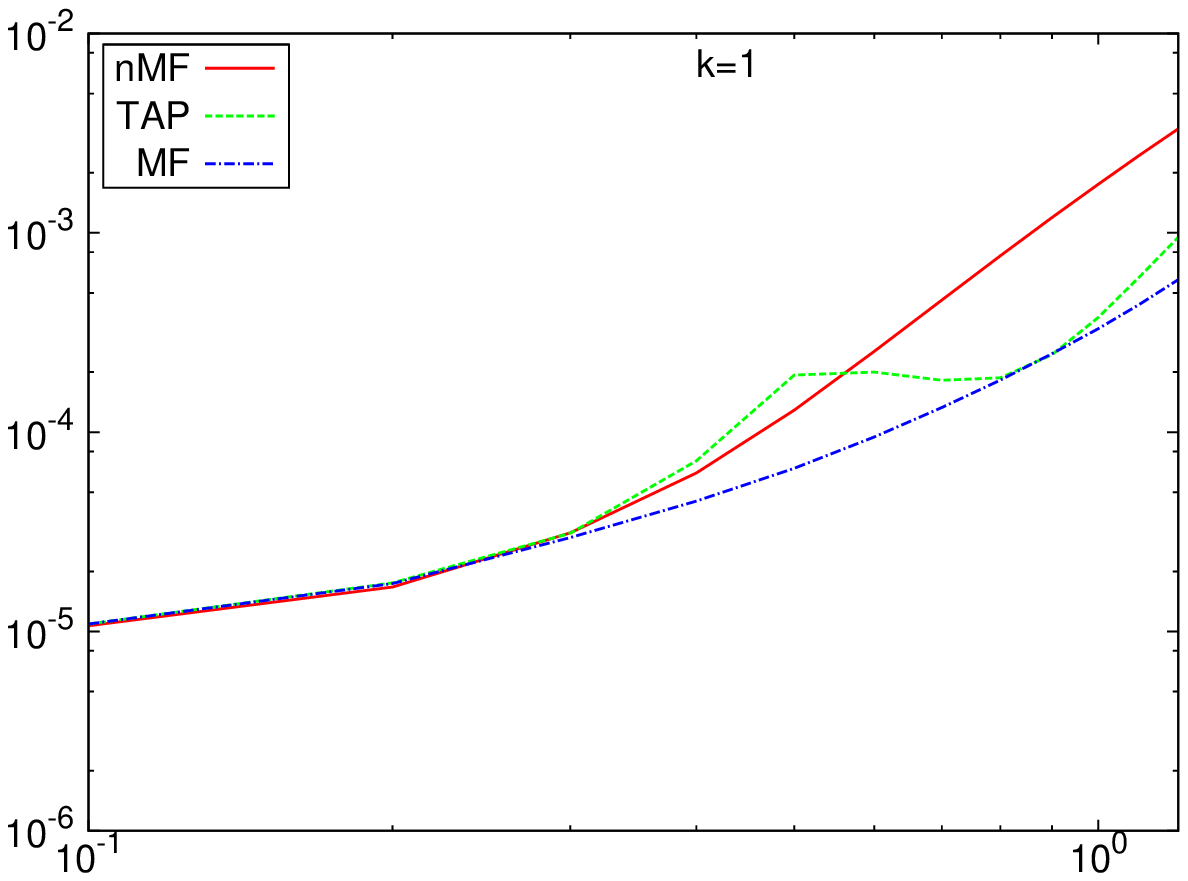}}
   \caption{

LEFT PANELS: Mean squarre error of the three methods on the infered couplings 
     $\overline{( J_{ij}^{\text{infered}} - J_{ij}^{\text{real}})^2}$ as a function of  
     $g$ for systems of size $N=100$ with zero external field, given $P=100000$ patterns, 
     averaged over $10$ realizations of the couplings. The different figures correspond 
     to different values of $k$. From top to bottom $k=0,0.25,0.5,0.75,1$.
     RIGHT PANELS: The same but with the addition of a sinusoidal external 
     field of period $10$ time steps and amplitude $0.5$.
  }  
  \label{fig:figInverse}
\end{figure}

The situation for the inverse problem is slightly more complicated. This is because, 
for strong couplings, the cubic equation that $F_i$ solves develops complex roots. 
In this case one can take three approaches: (i) take the nMF result, (ii) take 
the real part of the solution, (iii) take the solution for the largest $g$ for which the solutions are real. 
This value can be shown to be $F_i=1/3$. The results for the last two strategies almost coincide, 
with strategy (iii) performing slightly better in lower temperatures, so
we chose this one. In strategy (i) the results just coincide with the
nMF  approach after the temperature at which the cubic equation
for $F_i$ develops complex roots. The results from strategy (iii) are 
shown in Fig. \ref{fig:figInverse}. It is clear from this figure that 
nMF always performs worse than the other two and that the difference between 
the three methods vanishes in the high-temperature limit. 
On the other hand,  MF is superior, as expected, when one gets closer
to the asymmetric case i.e. for $k$ is close to 1.
The TAP result  has a more complicated behavior, due to the
intrinsic limitations imposed by the lack of real solutions of the
cubic equation at  strong couplings. However, one can notice that,
when $k$ is close to zero, there is a range of couplings $g$ where TAP becomes better than MF as it is expected. 

As can be seen in the right column of Fig. \ref{fig:figInverse}, the mean squared error 
$\overline{( J_{ij}^{\text{infered}} - J_{ij}^{\text{real}})^2}$ becomes larger for non-zero external fields.
This is a general feature of all three methods. Large fields and/or couplings are estimated with 
larger errors than small ones. This is because errors in the estimation 
of the empirical magnetizations/correlations, when the later  are close to $\pm 1$, produce large errors in the 
estimation of the fields/couplings (consider for example, in zeroth order approximation, a sigmoid map between $m_i$ and $h_i$ , 
and $c_{ij}$ and $J_{ij}$). Numerical simulations show that, for large external field amplitude, these errors become so important that the differences between the three methods are insignificant.

\section{Conclusions}

Within the mean field approaches that we have studied, the solution of
the inverse problem derives from the solution of the direct
problem. We have studied here three methods that provide an
approximate solution to the direct problem in the case of systems with
infinite range interactions. We have explored their behaviors 
on both the direct and the inverse problem in the case of SK models
with different degrees of symmetry of the interactions. As expected,
the MF approach is the best one when the degree of asymmetry is large
enough, but the TAP approach turns out to be slightly better in some
range of coupling strength when the couplings are more symmetric.
The nMF approach is just a first order approximation to both MF and TAP 
and is systematically worse than the other two methods.

As noted before, the derivation of inverse nMF and TAP rely on expanding
the $\tanh$ in the around the solutions of the nMF and TAP. This expansion
is not required for the MF solution: in the case with the
assumption of full asymmetry, the joint distribution of the local field 
to each pair of spins will be Gaussian and can be easily calculated. It is therefore
possible to write an exact equation relating $D_{ij}$ to $C_{ij}$ and the couplings
which in the limit of small $C_{ij}$ can be linearized and takes the form of Eq.\ \ref{DAJC}.
It would be interesting to see if a similar approach can be done within
the TAP framework: calculate the joint distribution of the local fields in a
systematic small coupling expansion, and use the
same procedure done in MF to relate $D_{ij}$ to $C_{ij}$.

In real applications, for instance in neural data analysis or gene regulation
network reconstruction, one does not deal with data generated from a 
model with the particular size dependence of the couplings of the SK model.
Our previous work shows that TAP and nMF perform at the same level in identifying 
the connections of a simulated neural network, and they both perform worse than 
the exact iterative Boltzmann like learning rule that one can write down for
the dynamical SK model \cite{Roudi11,HertzCNS10}. We will leave the comparison of
TAP, MF and the exact learning on biological data to future work.

\section*{Acknowledgement}
The work of MM and JS has been supported in part by the EC grant 'STAMINA', No 265496.

\end{document}